\documentclass[a4paper,11pt]{article}
\usepackage{jheppub}
\usepackage{dsfont}
\usepackage{amsmath,amssymb,amscd,amsfonts,mathtools}
\usepackage{slashed}
\usepackage{hyperref}
\usepackage[percent]{overpic}
\usepackage{xcolor}
\usepackage[utf8]{inputenc}
\usepackage{enumitem}
\title{Massless Fermions on a half-space:\\
The curious case of 2+1-dimensions}

\author{Shovon Biswas}
\author{and Gordon W. Semenoff}

\affiliation{Department of Physics and Astronomy, University of British Columbia\\
6224 Agricultural Road, Vancouver, British Columbia, Canada V6T 1Z1}

\emailAdd{shovon@phas.ubc.ca}
\emailAdd{gordonws@phas.ubc.ca}

\abstract{Boundary conditions for a massless Dirac fermion in 2+1 dimensions where the space is a half-plane are discussed in detail.  It is argued that linear boundary conditions that leave the Hamiltonian Hermitian generically break $C$ $P$ and $T$ symmetries as well as Lorentz and conformal symmetry.  We show that there is essentially one special case where a single species of fermion has  $CPT$ and the full Poincare and conformal symmetry of the boundary.   We show that, with doubled fermions, there is a second special case which respects $CPT$  but still violates Lorentz and conformal symmetry.  This second special case is  essentially the unique boundary condition where the Dirac operator has fermion zero mode edge states.  We discuss how the edge states lead to exotic representations of scale, phase and translation symmetries and  how imposing a symmetry requirement leads to edge ferromagnetism of the system.  We prove that  the exotic ferromagnetic representations are indeed carried by the ground states of the system perturbed by a class of interaction Hamiltonians which includes the non-relativistic Coulomb interaction.}

\makeatletter
\gdef\@fpheader{}
\makeatother

\begin{document} 
\maketitle
\flushbottom
 \section{Introduction}

One of the fascinating properties of the Dirac theory of fermions quantized in $2+1$ spacetime dimensions is that the mass term in the Dirac equation can, an in some cases must, break some of the discrete spacetime symmetries.  For example, for a single  two-component Dirac spinor, 
the Dirac equation with a mass term,
$
\left[\slashed \partial +m\right]\psi(x)=0
$,
while having the full Poincare symmetry expected of a relativistic wave equation, 
necessarily violates both spacetime parity ($P$) and time reversal ($T$) invariance, although it does have good charge conjugation ($C$) symmetry, $PT$ and  $CPT$.  

Of course the $P$ and $T$ symmetries can be restored by doubling the fermions so that the two species of fermions transform into each other under $P$ or $T$, but then the signs of the mass term for each of the two species must be different and the mass terms break a global $U(2)$ symmetry (or $O(2)$ symmetry in the case of Majorana fermions) which would rotate the two species into each other. One has the statement that, even for the case of doubled fermions, they can have a mass gap, but that gap must necessarily violate either $P$ and $T$ or an internal  $U(2)$ (or $O(2)$) symmetry as $U(2)\to U(1)\times U(1)$ or $O(2)\to Z_2\times Z_2$. 

In this note we wish to point out another fact, that even in the absence of mass, a single species of Dirac fermion defined on a 2+1-dimensional spacetime with a 1+1-dimensional boundary also violates
the discrete spacetime symmetries $P$ and $T$  through its boundary conditions. Like a mass term, it preserves $PT$.   In addition, like for a massive fermion, $P$ and $T$ can be restored by doubling the fermions, but then the boundary condition will violate a $U(2)$  symmetry.  Note that we do not include the Majorana option here, since, generically, there isn't one.  Unlike a mass term, boundary conditions for a single massless fermion on a half-space generally violate $C$.   Here, and in the rest of this paper, we will always take the boundary to be as symmetric as possible, specifically, if $x^\mu=(x^0,x^1,x^2)$, are the Cartesian coordinates of open, infinite Minkowski space, we will consider the half-space $x^1>0$. 

We will show that there is only one essentially unique special case of a boundary condition where a single species of fermion can have $C$ symmetry.   That theory, even  though it violates $P$ and $T$,  has $C$ and $CPT$ invariance and it also preserves the Poincare and conformal symmetry of the boundary.   It has the only boundary condition where a single species fermion can be a boundary conformal field theory.  In all other cases,  a single fermion species violates $C$ and $CPT$ and, coincidentally, it also violates the Lorentz and conformal symmetries of the boundary.  

We will also find that when the fermions are doubled, this special, highly symmetric theory has $C$, $P$ and $T$ symmetries as well as the full conformal symmetry of the boundary and, amongst all of the possibilities we consider, it is still the unique one which is a boundary conformal field theory. It is the theory described in equations (\ref{dirac_equations_11})-(\ref{dirac_equations_12}) of section \ref{doubled} below.  We will refer to it as the BCFT.  For doubled fermions, we shall call the two species ``the valleys'', after the terminology used in graphene. This boundary conformal field theory is the low energy theory that models clean graphene near charge neutrality and with an armchair boundary \cite{Dutta2010,Wang2021}.   

There is, for doubled fermions, a second  boundary condition that preserves $C$ as well as $P$ and $T$, distinct from the boundary conformal field theory described above. The difference is that its $C$ transformation, like $P$ and $T$, interchanges the two valleys, whereas in the BCFT it does not. Having $C$, $P$ and $T$ separately, it has $CPT$ symmetry but it is not Lorentz or conformal invariant.  The Lorentz and conformal symmetries are broken by the boundary condition and they cannot be restored by doubling.   This theory still has the residual translation symmetries and the scale symmetry of the boundary and it is an example of a scale invariant but not conformally invariant field theory.  It is the one described in equations  (\ref{dirac_equations_zig-zag})-(\ref{dirac_equations_zig-zag_1}) of section \ref{doubled} and we shall refer to it as the ``zig-zag theory'' as it describes the low energy limit of electrons near the charge neutral point in clean graphene with a zig-zag edge. The interesting feature of the zig-zag theory is that it has the unique boundary condition where the Dirac Hamiltonian can have static fermion zero mode edge states. The existence of bound states in a scale invariant theory where there are no dimensionful parameters is already unusual. In addition to that, fermion zero modes have long been associated with exotic  realizations of symmetries \cite{ Jackiw1976, Niemi1986} and this  zig-zag theory is no exception.  In this case it is the representation of the scaling symmetry which we shall find interesting. 

We will show that the only populations of the zero mode edge states with fermions which can be scale invariant are those where complete bands of the states are either completely filled or completely empty.  This fact, together with a requirement of  $U(1)$ charge neutrality will lead us to the valley ferromagnetic states.  Of course such states are only one or two of the enormous number of candidates amongst the degenerate ground states in a system with an infinite number of fermion zero modes and, in the free fermion theory, there is little besides their special symmetry property that distinguishes them.  To improve on this situation, we generalize an old argument, related to the Hund rules of atomic physics, and previously applied to the quantum Hall ferromagnetism that is seen in graphene \cite{semenoff2011} to prove that a perturbation of the free fermion theory by an instantaneous Coulomb interaction actually favours these ferromagnetic states -- in fact they are the exact ground state if the Coulomb interaction is weak enough. This establishes their importance and therefore the importance of scale invariant states.  

Practically any discussion of massless relativistic fermions in 2+1-dimensions should have applications to Dirac materials \cite{Wehling2014}, examples of which are
d-wave superconductors, graphene, and topological insulators. 
Graphene has long been a venue where field theoretical ideas about 2+1 dimensional Dirac fermions could find application \cite{Semenoff1984} and boundary conditions for the Dirac field corresponding to various boundaries have been found \cite{Akhmerov2008,Ostaay2011} and are compatible with what we shall find in this paper. Moreover, the occurrence of edge states on zig-zag edges of graphene \cite{Fujita1996,Nakada1996} was discussed  years before graphene was readily available in the laboratory and their fascinating properties have inspired literally thousands of research papers. The edge states themselves are readily detected by experimental techniques such as scanning tunnelling microscopy\cite{Niimi2006,Kobayashi2006}, high resolution transmission electron microscopy \cite{Liu2009} and atom-by-atom spectroscopy \cite{Suenaga2010}.   Most theoretical modelling using a variety of techniques, mean field theory\cite{Fujita1996,Jung2009,Jung2009a}, density functional theory\cite{Son2006,Son2006a}, numerics \cite{Hikihara2003,Dutta2008,Feldner2010} and perturbation theory \cite{Karimi2012,Shi2017}, indicate that the edge states favour taking up a ferromagnetic configuration, although this has yet to be seen reliably in experiments.  Such an edge ferromagnetic  state is expected to have important technological applications in spintronics \cite{Son2006,Son2006a}.  The theoretical computations normally use the lattice tight binding model of graphene and various types of interactions and mean field theory arguments.  
Our results are confined to the scale invariant continuum theory that would describe the electrons in a low energy, large wavelength limit of clean graphene near charge neutrality.  In that context, our proof that the scale invariant valley or spin ferromagnetic states are actually the true ground states of such a system with a weak Coulomb interaction added  is in line with other results in this vein.    Affleck \cite{Karimi2012,Shi2017} et.al.~give a strong argument that the ground state of charge neutral graphene with a weak Hubbard interaction is edge ferromagnetic and they also argue, with somewhat less rigour that the same should apply with the Coulomb interaction.  Our work uses a similar technique,  is in agreement with them and, in addition,  it solves the issue of edge-bulk hybridization which they left as an open problem. 
It also treats the Coulomb interaction directly and it is easily generalized to other interactions. 
Finally, we emphasize the tantalizing possibility that this edge state ferromagnetism could be confirmed by experiment in the near future.

One interesting property of the valley ferromagnetic states of the system with two species (two valleys and one spin polarization)  that we find, and which we would like to understand better, is an induced momentum density for
momentum parallel to the boundary (see equation (\ref{induced_momentum}),
$$ 
\left< {\bf T}^{02}(x) \right>    =\pm\frac{1}{4\pi}\frac{1}{x_1^3} 
$$
where $ {\bf T}^{\mu\nu}(x)$ is the stress tensor and the sign on the right-hand-side depends on the polarization of the valley ferromagnet.
This equation is relevant to spin polarized graphene.  With two degenerate spin states, it can easily cancel between two spin degrees of freedom. 
It can also easily be shown that
this apparent buildup of momentum density is not due to charge transport as the  current vanishes, $$\left< {J}^{2}(x) \right>=0$$

The remainder of this paper is organized as follows.  In section \ref{one_valley}, we will discuss the possible boundary conditions for the Dirac field.  
This includes a discussion of both the discrete and continuous spacetime symmetries.  
It is where we show that there is essentially one unique Lorentz and conformal invariant possibility.  
In section \ref{doubled}, we look at the possibility of restoring the discrete spacetime symmetries by doubling the number of fermions.  We find and then we concentrate on two interesting cases, which we will call the BCFT and the zigzag theories. In section \ref{edge}, we explore the zigzag theory in more detail by solving the Dirac equation explicitly and then discussing the possible ground states of the quantum field theory.  It is there that we argue that scale invariance is closely tied to valley ferromagnetism. 
In section \ref{resolution}, we prove that the ferromagnetic states are the ground state of the zigzag theory deformed by a weak nonrelativistic Coulomb interaction. 
In section \ref{conclusions}, we offer conclusions.

\section{Single species of massless Dirac Fermion}\label{one_valley}

Consider a single Dirac Fermion which obeys the massless Dirac equation
\begin{align}\label{Dirac_equation}
i\gamma^\mu\partial_\mu \psi(x)=0
\end{align}
with the Dirac field obeying equal time anti-commutation relations
\begin{align}\label{etacr}
& \left\{\psi(x),\psi^{\dagger}(x')\right\}\delta(x^0-{x'}^0)=\delta(x-x') 
\end{align}
and 
defined   on the half-space $(x_0,x_1,x_2)$ with $x_1>0$.  The spinor field must also obey a boundary condition which we will assume is linear and we will take as
\begin{align}\label{boundary_condition} 
\lim_{x^1\to 0} \left( 1-\Pi\right)\psi(x)=0
\end{align}
where $\Pi$ is a $2\times2$ matrix.  In the following we will discuss the possible forms of $\Pi$. Note that the anti-commutator (\ref{etacr}) is valid only in the region where $x^1,{x'}^1>0$. 

We will consider linear boundary conditions where   the Dirac Hamiltonian 
\begin{align}\label{Dirac_Hamiltonian}
H=i \gamma^0\vec\gamma\cdot\vec\nabla
\end{align}
defined on the space with a boundary is 
 a self-adjoint differential operator. 
 A necessary condition  is that the normal component of the  fermion particle flux vanishes there, that is,
\begin{align}\label{current_boundary}
\left.
-\psi^\dagger(x)\gamma^0\gamma^1\psi(x)\right|_{x_1=0}=0 
\end{align}
 This requirement is satisfied by a two-parameter family of local, linear boundary conditions of the sort
in equation (\ref{boundary_condition}) where  
\begin{align} \label{Pi}
\Pi= i\cosh \chi \cos\theta\gamma_0+\cosh\chi\sin\theta\gamma_1+i\sinh\chi\gamma^2
\end{align}  
Here the parameter $\chi$ can be any real number and $\theta\in[-\pi,\pi]$.
A constraint on $\Pi$ that is solved by the ansatz (\ref{Pi}) is simply that $(1-\Pi)$ must be proportional to a projection operator with rank one, so that, at the boundary, it constrains one of the two degrees of freedom in the Dirac spinor $\psi(x)$. 
In the next section, we shall see that the boundary condition  (\ref{boundary_condition})  with $\Pi$ given in (\ref{Pi}) generally violates both discrete
and continuous spacetime symmetries.  Then we will focus on a few special cases which can be more symmetric.
  
\subsection{Discrete spacetime symmetries}

To examine the discrete spacetime symmetries of the Dirac equation, it is useful to observe that the Dirac matrices are $2\times 2$ matrices which have a Majorana representation in which they are real,
\begin{align}\label{majorana_representation}
\gamma_0=-i\sigma_2~,~~\gamma_1=\sigma_1~,~~\gamma_2=\sigma_3~~,~~\gamma_\mu=\gamma_\mu^*
\end{align}
where 
\begin{align}\label{Pauli_matrices}
\sigma_1=\left[\begin{matrix}0&1\cr 1&0\cr \end{matrix}\right],~
\sigma_2=\left[\begin{matrix}0&-i\cr i&0\cr \end{matrix}\right],~
\sigma_3=\left[\begin{matrix}1&0\cr 0&-1\cr \end{matrix}\right]
\end{align}
 are the Pauli matrices. 
These matrices have the algebraic property
 \begin{align}
 \gamma^\mu\gamma^\nu = \epsilon^{\mu\nu\lambda}\gamma_\lambda + \eta^{\mu\nu}
  \end{align}
 where $\epsilon^{\mu\nu\lambda}$ is the totally anti-symmetric tensor with 
 $\epsilon^{012}=1$ and $\eta^{\mu\nu}={\rm diag}(-1,1,1)$ is the metric of Minkowski spacetime.
 
 When the Dirac matrices are in the Majorana representation (\ref{majorana_representation}) the discrete spacetime symmetries, $C$, $P$ and $T$ of the Dirac equation are given by the transformations
  \begin{align}
 &       
C:\psi(x) =\psi^*(x)  &{\rm when}~\gamma_\mu=\gamma_\mu^*\label{C}
\\
&P:\psi(x) =i \gamma^2\psi(x')
 & x'= (x_0,-x_1,x_2)\label{P}
\\
&T:\psi(x) =\mathcal Ti  \gamma^0\psi(\tilde x)
 &\tilde x=(-x_0,x_1,x_2)\label{T}
\end{align}
In equation (\ref{T}), $\mathcal T$ is the complex conjugation operator.  Here we have adjusted the phases of the $P$ and $T$ transformation so that, when acting on fermions, $P^2=-1$, $T^2=-1$ and $(CPT)^2=-1$, although this is not important for the current discussion. 

 There is no choice of the parameters $\chi$ and $\theta$ for which the boundary condition (which we recopy here for convenience)
 \begin{align}\label{boundary_condition_1}
&\lim_{x^1\to 0} \left( 1-\Pi\right)\psi(x)=0\\
&\Pi= i\cosh \chi \cos\theta\gamma_0+\cosh\chi\sin\theta\gamma_1+i\sinh\chi\gamma^2
\label{boundary_condition_2}
\end{align}
is compatible with either the $P$ or the $T$ transformation.  All possible choices of the linear boundary condition
break both $P$ symmetry  and $T$ symmetry.  However, all of the possible choices preserve the combined $PT$ symmetry. 

Since $PT$ is a good symmetry, we could ask about $CPT$ which would require invariance under $C$. 
The $C$ transformation is simple in the Majorana representation where the Dirac operator $\slashed\partial$ is real.  
Invariance under $C$ further requires that the boundary condition is real.   Of the family of conditions in equations (\ref{boundary_condition_1}) and (\ref{boundary_condition_2}), there are only two essentially  equivalent boundary conditions that preserve  $C$ and therefore $CPT$.  They are gotten from the choice of parameters
$\chi=0$ and $\theta=\pm\frac{\pi }{2}$ in $\Pi$ above so that 
 \begin{align}\label{bcft}
  \Pi=\pm\gamma_1~~,~~\lim_{x^1\to 0} \left( 1\mp \gamma^1\right)\psi(x)=0
 \end{align}
 Coincidentally, we will see shortly that these are also the unique choice of boundary condition that 
 preserves the Lorentz and conformal symmetry of the boundary. It is interesting to find such a strong 
 correlation between CPT and Lorentz invariance in the context of a boundary field theory.  The Dirac theory on a half-space with one of the boundary conditions in equation (\ref{bcft}) is the starting point for construction of a boundary conformal field theory with fermion content. 
 
 Another special boundary condition that will be of interest to us is one which allows edge states.  These are zero energy eigenstates of the Dirac Hamiltonian
 (\ref{Dirac_Hamiltonian}) which are localized at the boundary.  In the family of boundary conditions in equations (\ref{boundary_condition_1}) and (\ref{boundary_condition_2}) there is an essentially unique choice for which this can occur, the choices $\chi=0$ and $\theta=\pm\pi$, 
  \begin{align}\label{zig-zag_Pi}
  \Pi=\pm i\gamma^0~~,~~\lim_{x^1\to 0} \left( 1\pm i\gamma^0\right)\psi(x)=0
  \end{align}
  As we have already discussed, this boundary condition violates all of the $C$, $P$, $T$ and $CPT$ symmetries and we will see shortly that it also violates the Lorentz and conformal symmetry of the boundary.   On the other hand, the edge states have interesting implications for the physical properties of such a system.  A discussion of them and their properties will occupy most of the bulk of this paper.
      
 \subsection{Continuous Spacetime symmetries}

 A continuous space-time symmetry is symmetry under an infinitesimal coordinate transformation $x^\mu\to x^\mu+f^\mu(x)$ where the spinor field transforms as
\begin{align}
\delta_f \psi(x) =\left[ f^\mu(x)\partial_\mu +\frac{1}{4}\epsilon^{\mu\nu\lambda}\partial_\mu f_\nu(x)\gamma_\lambda + \frac{1}{3}\partial_\mu f^\mu(x)\right]~\psi(x)
\end{align}
On infinite $2+1$-dimensional spacetime  -- without the boundary -- the above transformation is a symmetry of the Dirac equation (\ref{Dirac_equation})
and the symplectic structure when $f^\mu(x)$ is a conformal Killing vector of three-dimensional Minkowski space.  A conformal Killing vector obeys the conformal Killing equation
\begin{align}
\partial_\mu f_\nu(x)+\partial_\nu f_\mu(x)-\frac{2}{3}\eta_{\mu\nu}\partial_\lambda f^\lambda(x)=0
\end{align}
The ten solutions of this equation are 
\begin{align}
f^\lambda(x) = \left\{ a^\lambda,~ \omega^\lambda_{~\rho}x^\rho, ~c x^\lambda, ~x^\lambda b\cdot x-b^\lambda x^2/2\right\}
\label{killing_vectors}
\end{align}
corresponding to translations $\delta x^\lambda=a^\lambda$, Lorentz transformations    $
\delta x^\lambda=\omega^\lambda_{~\rho}x^\rho$ where $\omega_{\mu\nu}=-\omega_{\nu\mu}$ , dilatations $\delta x^\lambda=cx^\lambda$   and conformal transformations $\delta x^\lambda = x^\lambda b_\nu x^\nu-b^\lambda x_\nu x^\nu/2$ with parameters $b^\lambda$, respectively.  

When the boundary is present, these symmetries are reduced to those which preserve the geometry of the boundary and which also preserve the boundary condition. The symmetries which preserve the geometry of the boundary are just the symmetry transformations of
the boundary itself which in our case is the two-dimensional Minkowski spacetime with coordinates $x^\mu=(x^0,0,x^2)$.  
The remaining symmetries are then time translations and space translations along the boundary, $a^\mu=(a^0,0,a^2)$, the  Lorentz transformation along the boundary, with the only nonzero components of $\omega^{\lambda}_{~\rho}$ being $   \omega_{20}=  -\omega_{02}$,   the dilatation symmetry with parameter $c$ and $\delta x^\mu=c(x^0,x^1,x^2)$ and the conformal transformations with $b^\mu=(b^0,0,b^2)$. 
These transformations are generated by Noether currents related to the stress tensor  and Noether charges which generate the transformations, respectively
\begin{align}
   &-i\left[P_\mu,~\Psi(x)\right]=\partial_\mu \psi(x) ~,~~\mu=0,2 \\
 &-i\left[  M,~\Psi (x)\right]= \left( x_2\partial_ 0-x_0\partial_2+\frac{1}{2}\gamma^1\right) \psi (x)&  \label{Lorentz}\\
&-i \left[  \Delta,~\Psi (x)\right]=  \left(x^\mu \partial_\mu +1\right)~ \psi(x)&  \\
&-i\left[   {K_\lambda},~\Psi(x)\right]=  \bigg\{\left(x_\lambda x^\mu-\frac{x^2 }{2} \delta_\lambda^\mu\right)\partial_\mu + \frac{1}{2} \epsilon_{\lambda\rho\sigma} x^\rho\gamma^\sigma +x_\lambda\biggr\} \psi(x)
,~\lambda=0,2
 \label{Conformal}
\end{align}
Because of the action of the spin operator in the transformation, for generic values of $\chi$ and $\theta$, the boundary conditions in equations (\ref{boundary_condition_1}) and (\ref{boundary_condition_2}) 
are not invariant under the Lorentz transformation generated by $M$ 
in equation (\ref{Lorentz}), nor  are they invariant under the conformal transformations generated by $K_0$ or $K_2$ in equation (\ref{Conformal}). The transformation from the spin operators in both of those cases would not be compatible with the boundary conditions.
The generic boundary condition therefore reduces the symmetry of the theory  from the conformal symmetry of the boundary to a subset consisting of time translations, 
spatial translations along the boundary and dilatations.
  These symmetries are governed  by the sub-algebra.generated by $P_0$, $P_2$ and $\Delta$ of the full conformal algebra.  

There is one special case of the boundary conditions which is compatible with the spin part of the transformation in $M,~K^0$ and $K^2$. It is the case where $\chi=0$, $\theta=\pm\frac{\pi}{2}$ and $\Pi=\pm\gamma^1$,  where the boundary condition preserves the Lorentz and conformal invariance.   As we have discussed earlier in this section, this is also the unique special case where the boundary condition preserves $C$ and $CPT$ symmetries. The single relativistic fermion theory in that special case is a boundary conformal field theory and as such it is interesting in its own right.  It would be very interesting to study it further using conformal techniques.  We do not have time or space to do it justice here and we put it off for future work.  Instead we will concentrate on another special case. 
\section{Doubled fermions}\label{doubled}

In the previous section, we learned that $P$ and $T$ can never be good symmetries of a single massless Dirac theory for any acceptable linear boundary condition. 
In this section, we will double the degrees of freedom so that there are two species of fermions, $\psi^{(1)}(x)$ and $\psi^{(2)}(x)$ satisfying identical massless Dirac equations
\begin{align}
i\gamma^\mu\partial_\mu \psi^{(1)}(x)=0~~,~~i\gamma^\mu\partial_\mu \psi^{(2)}(x)=0 
\end{align}
and, as is common in the literature on Dirac materials like graphene, we will call the two species the two ``valleys''.\footnote{We caution the reader that the valleys that we refer to can differ from the specific species that are commonly referred to as valleys in graphene which are the individual sets of electronic excitations near the two $K$-points.  In our continuum approach we do not have information about the identity of the $K$ points.  Instead we will chose a basis for our two species of fermions which diagonalize the boundary conditions in equation (\ref{tbc1}) and  (\ref{tbc2}). A more general choice would have a Hermitian matrix with eigenvalues $\pm1$ acting on the valley indices on the right-hand-side of those equations. Our ``valley'' fermions are generally linear combinations of the usual graphene valley electrons for the (BCFT) armchair edge, but they coincide with the usual definition for the zig-zag edge. }
   Now, we can do one of two things.  We could demand that each valley has the same boundary condition.  Then we would simply have two identical copies of the $P$ and $T$ violating fermions that we studied in the previous section.  The system would have a ``valley symmetry'' under the transformation
$\psi^{(A)}(x)\to U^A_{~B}\psi^{(B)}(x)$ where $U^A_{~B}$ is a  $2\times2$ unitary matrix.

 Alternatively, 
 we could allow the valleys to transform into each other under $P$ and $T$.  Cursory examination
of the possibilities shows us that the following system is $P$ and $T$ symmetric,
\begin{align}\label{dirac_equations}
&i\gamma^\mu\partial_\mu \psi^{(1)}(x)=0~~,~~i\gamma^\mu\partial_\mu \psi^{(2)}(x)=0 \\
& \left\{\psi^{(A)}(x),\psi^{(B)\dagger}(x')\right\}\delta(x^0-{x'}^0)=\delta^{AB}\delta(x-x') \\
&  \psi^{(1)}(x)_{x^1\to 0} =  ( i\cosh \chi \cos\theta\gamma_0+\cosh\chi\sin\theta\gamma_1+i\sinh\chi\gamma_2) \psi^{(1)}(x)_{x^1\to 0}
\label{tbc1}  \\
 & \psi^{(2)}(x)_{x^1\to 0}= 
 (-i\cosh \chi \cos\theta\gamma_0 - \cosh\chi\sin\theta\gamma_1+ i\sinh\chi\gamma_2)\psi^{(2)}(x)_{x^1\to 0} \label{tbc2} 
 \end{align}
 where $T$ and $P$ are
 \begin{align}
&T:\left\{ \begin{matrix} \psi^{(1)}(x)~\to~\mathcal Ti \gamma^0\psi^{(2)}(\tilde x) \cr
\psi^{(2)}(x)~\to~\mathcal T i\gamma^0\psi^{(1)}(\tilde x) \cr \end{matrix}\right. \label{restored_p}\\ &
P:\left\{ \begin{matrix} \psi^{(1)}(x)~\to~ i\gamma^2\psi^{(2)}(  x') \cr
\psi^{(2)}(x)~\to~ i\gamma^2\psi^{(1)}(  x') \cr \end{matrix}\right.   \label{restored_t}
\end{align}
for any value of the parameters $\chi$ and $\theta$.  We thus have a large family of boundary conditions which are compatible with $P$ and $T$ symmetry.  
We note that, on the other hand, for any values of $\chi$ and $\theta$, the boundary condition  breaks the valley symmetry -- reducing $U(2)$ to $U(1)\times U(1)$. 
  
Now that, in equations (\ref{dirac_equations})-(\ref{restored_t}), we have restored the $P$ and $T$ symmetries, 
we can again ask whether any of the family of boundary conditions there is also consistent with imposing a $C$ symmetry.
For this, there are only two possibilities.  In the first possibility, we assume that $C$ does not interchange the valleys, that is $(1)\to(1)$ and $(2)\to(2)$. Then
it can be so only if $\chi=0$ and $\theta=\pm\pi/2$ and  (\ref{dirac_equations})-(\ref{restored_t}) become
\begin{align}\label{dirac_equations_11}
&i\gamma^\mu\partial_\mu \psi^{(1)}(x)=0~~,~~i\gamma^\mu\partial_\mu \psi^{(2)}(x)=0 \\
& \left\{\psi^{(A)}(x),\psi^{(B)\dagger}(x')\right\}\delta(x^0-{x'}^0)=\delta^{AB}\delta(x-x') \\
&  \left[1- \gamma_1 \right] \psi^{(1)}(x)_{x^1\to 0}=0  
 ~,~\left[1 + \gamma_1\right] \psi^{(2)}(x)_{x^1\to 0}=0\label{tbc22} 
 \end{align}
 where 
 \begin{align}
&T:\left\{ \begin{matrix} \psi^{(1)}(x)\to\mathcal T i\gamma^0\psi^{(2)}(\tilde x) \cr
\psi^{(2)}(x)~\to~\mathcal T i \gamma^0\psi^{(1)}(\tilde x) \cr \end{matrix}\right. ~\\ &
P:\left\{ \begin{matrix} \psi^{(1)}(x)\to i\gamma^2\psi^{(2)}(  x') \cr
\psi^{(2)}(x)\to i\gamma^2\psi^{(1)}(  x') \cr \end{matrix}\right.   \\ &
C:\left\{ \begin{matrix} \psi^{(1)}(x)\to\psi^{(1)*}(  x)\cr
\psi^{(2)}(x)\to\psi^{(2)^*}(  x)\cr \end{matrix}\right.  
\label{dirac_equations_12}
\end{align}
The above equations (\ref{dirac_equations_11})-(\ref{dirac_equations_12}) now describe  a $C$, $P$ and $T$ invariant boundary conformal field theory which we refer to as BCFT. This theory  describes the very low energy limit of clean, charge neutral (spin polarized) graphene with an armchair edge.  

The other possibility for obtaining a $C$ invariant theory is to let the $C$ transformation interchange the valleys $(1)\leftrightarrow(2)$. There is also a unique choice in equations  (\ref{dirac_equations})-(\ref{restored_t}),  which obtains $C$ invariance with this interchange, where the parameters are $\chi=0$ and $\theta=\pm\pi$, and  (\ref{dirac_equations})-(\ref{restored_t}) become
\begin{align}\label{dirac_equations_zig-zag}
&i\gamma^\mu\partial_\mu \psi^{(1)}(x)=0~~,~~i\gamma^\mu\partial_\mu \psi^{(2)}(x)=0 \\
& \left\{\psi^{(A)}(x),\psi^{(B)\dagger}(x')\right\}\delta(x^0-{x'}^0)=\delta^{AB}\delta(x-x') \\
&  \left[1+i \gamma_0 \right] \psi^{(1)}(x)_{x^1\to 0}=0~,~
  \left[1 - i\gamma_0\right] \psi^{(2)}(x)_{x^1\to 0}=0\label{tbc23}
  \end{align}
  where
  \begin{align}
&T:\left\{ \begin{matrix} \psi^{(1)}(x)\to\mathcal Ti \gamma^0\psi^{(2)}(\tilde x) \cr
\psi^{(2)}(x)\to\mathcal T i\gamma^0\psi^{(1)}(\tilde x) \cr \end{matrix}\right. \\ &
P:\left\{ \begin{matrix} \psi^{(1)}(x)\to i\gamma^2\psi^{(2)}(  x') \cr
\psi^{(2)}(x)\to i\gamma^2\psi^{(1)}(  x') \cr \end{matrix}\right.   \\ &
C:\left\{ \begin{matrix} \psi^{(1)}(x)\to\psi^{(2)*}(  x)\cr
\psi^{(2)}(x)\to \psi^{(1)^*}(  x) \cr \end{matrix}\right.  
\label{dirac_equations_zig-zag_1}
\end{align}
This theory is $C$, $P$ and $T$ invariant.  We note and emphasize here that the doubling of the fermion degrees of freedom cannot restore a continuous space-time symmetry that is broken by boundary conditions.  The system described by
equations (\ref{dirac_equations_zig-zag})-(\ref{dirac_equations_zig-zag_1})  boundary conditions still violate the Lorentz and conformal symmetries of the boundary so that its only continuous spacetime symmetries are under time translation, space translation parallel to the boundary and scale transformations.  It is still, on the other hand, a doubled version of the unique boundary condition that allows the Dirac equation to have zero energy edge state solutions.  
A discussion of those solutions and their implications for the realization of the remaining symmetries will occupy the rest of this paper.  This theory  describes the very low energy limit of clean, charge neutral (spin polarized) graphene with an zig-zag edge, as well as a host of other possible edges that are related to the zig-zag \cite{Akhmerov2008,Ostaay2011}.

\section{Edge states}\label{edge}

Let us consider the zig-zag system described by the equations and boundary conditions in equations (\ref{dirac_equations_zig-zag})-(\ref{dirac_equations_zig-zag_1}). 
If we look for solutions of the Dirac equation of the form $\psi^{(A)}(x)=\left[\begin{matrix}u^{(A)}(x)\cr v^{(A)}(x)\cr\end{matrix}\right]$
the equation and boundary condition become  
 \begin{align}
&\left[ \begin{matrix}\partial_y & \partial_x-\partial_t \cr \partial_x+\partial_t & -\partial_y\cr\end{matrix}\right]
\left[ \begin{matrix} u^{(A)}(t,x,y)\cr v^{(A)}(t,x,y)\cr \end{matrix}\right]=0
\nonumber \\
&u^{(1)}(t,0,y)=-iv^{(1)}(t,0,y) 
~,~~u^{(2)}(t,0,y)=iv^{(2)}(t,0,y)\label{zigzag_valley_1}
\end{align}
Note that, in this Majorana representation of the Dirac matrices, the Dirac equation is real. However, the boundary conditions are not real, so the solutions which we shall find are complex. 

The equation and boundary conditions in   (\ref{zigzag_valley_1})  have the explicit solutions (with $(x^0,x^1,x^2)$ labeled $(t,x,y)$)
\begin{align}
&
\psi^{(1)}_{k,\ell\pm}(x)~=\frac{e^{-i\omega t+iky}}{\sqrt{ 8\pi\omega^2}}
\left[\begin{matrix}
(ik+\ell+\omega)e^{i\ell x} + (-ik+\ell-\omega)e^{-i\ell x} \cr
i(ik+\ell-\omega)e^{i\ell x} +i(-ik+\ell+\omega)e^{-i\ell x} \cr
\end{matrix}\right] 
\label{solutions_1} \\ 
&
\psi_{k,\ell\pm}^{(2)}(x)~=\frac{e^{-i\omega t+iky}}{\sqrt{ 8\pi\omega^2 }}
\left[\begin{matrix}(-ik+\ell+\omega)e^{i\ell x} + (ik+\ell-\omega)e^{-i\ell x} \cr
i(ik-\ell+\omega)e^{i\ell x} +i(-ik-\ell-\omega)e^{-i\ell x} \cr
\end{matrix}\right] 
\label{solutions_2}\\
& \ell\geq 0, ~-\infty<k<\infty, \nonumber \\
&\psi^{(A)}_{k,\ell+}(x)~{\rm has}~ \omega=\sqrt{k^2+\ell^2};~\psi^{(A)}_{k,\ell-}(x)~{\rm has}~ \omega=-\sqrt{k^2+\ell^2}
  \nonumber \\\label{solutions_3}
\\&\psi^{(1)}_{0,k}(x)~=~\sqrt{ \frac{|k|}{2{\pi}}}~e^{iky - kx}\left[ \begin{matrix} 1\cr i\cr \end{matrix}\right] ~~,~~k>0,~\omega=0
\label{solutions_4}\\
&\psi^{(2)}_{0,k}(x)~=~~\sqrt{ \frac{|k|}{2{\pi}}}~e^{iky + kx}\left[ \begin{matrix} 1\cr- i\cr \end{matrix}\right] ~~,~~k<0,~\omega=0
\label{solutions_5}
\end{align}
The above solutions are Dirac spinors $\psi^{(1)}_{k,\ell\pm}(x)$ and $\psi^{(2)}_{k,\ell\pm}(x)$ which propagate in the bulk of the half-space and we shall refer to them as ``bulk states''.  Their subscripts $\pm$ have two choices of sign, $+$ corresponds to a positive energy solution and $-$ corresponds to a negative energy solution. Then, there are also the spinors $\psi^{(1)}_{0,k}(x)$ and $\psi^{(2)}_{0,k}(x)$ which are localized near the edge of the half-plane at $x^1=0$ and we will refer to them  as ``edge states''. They have zero energy. 

We have normalized the wave-functions and it is easy to check that they obey the appropriate completeness relation -- which indicates that the entire set of bulk and edge states are required by the equal-time anti-commutation relation (\ref{etacr}).   The energy of each of the single particle bulk states is given by its value of $\omega$ and, for each valley $(1)$ and $(2)$, there is a band of positive energy bulk states and a band of negative energy bulk states.  As  well, the  infinite number of edge states all of which have zero energy form what is sometimes called a ``flat band''. 

   To find these solutions we have used the fact that the system is translation invariant along the $t=x^0$- and $y=x^2$-directions, the latter being parallel to the boundary.  The eigenvalues  of the energy and momentum 
operators which generates translations in the $t$- and $y$-directions are  the variables $\omega$ and $k$ which appear in  the solutions. It is interesting that the edge states in a given valley, besides having $\omega=0$,  all have momenta with the same sign, either $k>0$
in valley $(1)$ or $k<0$ in valley $(2)$.

We are interested in the states of this system near the Dirac vacuum where all of the negative energy states are occupied and all of the positive energy states are empty. This is clearly unambiguous for the bulk states where there is a clear distinction between positive and negative energy states.  We begin construction of our candidate ground states for this system by assuming that all of the  positive energy bulk states are empty and all of the negative energy bulk states are filled. There is an equivalent ``second quantized'' version of this statement -- that we choose the state that is the vacuum for both fermions and anti-fermions.

What is left to decide on is the distribution of fermions in the zero energy edge states.  There is clearly an enormous vacuum degeneracy since any such distribution of fermions into the zero modes has the same energy as any other, no matter what the number of particles  or with which values of momenta or which valleys the particles occupy.  In the following, we will concentrate on finding special occupations of the states which have certain symmetry properties, an important one being scale invariance.  Later on we will explain why we are particularly interested in scale invariant states. 

Let us ask what are the possible scale invariant states of the many-particle system.  A scale transformation, $x^\mu\to\Lambda x^\mu$ acts on the positive, negative and zero energy states  as the transformation
\begin{align}
  \psi_{k,\ell}^{(A)}(\  x) ~~\to~~\Lambda\psi^{(A)}_{k,\ell\pm}(\Lambda x)~=~ \Lambda\psi^{(A)}_{\Lambda k,\Lambda\ell\pm}(x)\\
\psi^{(A)}_{0,k}( x)   ~~\to~~ \Lambda\psi^{(A)}_{0,k}(\Lambda x)~=~  \Lambda^{\frac{1}{2}}\psi^{(A)}_{0,\Lambda k}(x)
\end{align}

An orbit of the scaling transformations acting on a function is the set of all functions that are generated from it by scale transformation with  $0<\Lambda<\infty$.
An orbit of a particular bulk state is the infinite set of  bulk states with all magnitudes but the same signs of $k$ and $\omega$ (remember that $\ell>0$ in the solutions
(\ref{solutions_1})-(\ref{solutions_3})),
$$
{\rm Orbit~of~scaling}~=~\{ \Lambda\psi^{(A)}_{\Lambda k,\Lambda\ell\pm }(x), \forall \Lambda\in (0,\infty) \}
$$
In each valley the set of all bulk states thus divides into four subsets according to the signs of $k$ and $\omega$, each subset being an orbit of scaling. 

On the other hand, as can be seen by inspecting (\ref{solutions_4})-(\ref{solutions_5}), the set of all of the edge states in a given valley comprise a single orbit of scaling, 
$$
{\rm Orbit~of~scaling}~=~\{ \Lambda^{\frac{1}{2}}\psi^{(A)}_{0,\Lambda k}(x), \forall \Lambda\in (0,\infty) \}
$$

A scale invariant state of the quantum field theory, that is, of this many-fermion system should have every orbit of the scale transformations either completely filled with fermions or completely empty.  For the bulk states in a given valley, we see that the negative energy states consist of two complete orbits, one with $k>0$ and one with $k<0$, as do the positive energy states.  Thus, completely filling the negative energy bulk states and leaving the positive energy bulk states completely empty is a scale invariant filling of the bulk states.  

  For occupations of the edge states, on the other hand, the only possibilities are to either completely fill or leave completely empty the zero mode states in a given valley. This leaves us with four possible scale invariant ground states.  In all four cases the positive energy bulk states are completely empty and negative energy bulk states are completely filled.  Then, either the edge states have both valleys  completely filled, or both valley's completely empty, or the two possibilities where one valley's edge states are completely filled and the other valley's edge states are completely empty. Scale symmetry alone does not tell us which of these four possibilities is the correct one.  For that, we need more information.

The Dirac equation and boundary conditions in equation (\ref{dirac_equations_zig-zag}) has two $U(1)$ symmetries, and two conserved currents, $\bar\psi^{(1)}(x)\gamma^\mu\psi^{(1)}(x)$ and $\bar\psi^{(2)}(x)\gamma^\mu\psi^{(2)}(x)$, one for each valley.   We will call the sum of the Noether charges corresponding to those two symmetries
\begin{align}\label{charge}
Q=\int d^2x \biggl\{ \frac{1}{2}\biggl[ \psi^{(1)\dagger}(x),\psi^{(1)}(x)\biggr]~+~  \frac{1}{2}\biggl[\psi^{(2)\dagger}(x) ,\psi^{(2)}(x)\biggr] \biggr\}
\end{align}
 the ``electric charge''.   In this vein, we will also assume that there is a background charge density which is $-1/2$ unit per degree of freedom, so that many fermion states with exactly half of the energy levels occupied are charge neutral.  The Dirac commutator prescription for ordering the operators in $Q$ in equation (\ref{charge}) is one way to take this neutralizing background charge density into account. As defined in (\ref{charge}) it is easy to confirm that $Q\to -Q$ under $C$ and that
it is invariant under $P$ and $T$. 
  
  Then, given the symmetry of the spectrum between positive and negative energy bulk states we would conclude that, to achieve overall charge neutrality of the many-fermion system when the bulk negative energy states are filled and positive energy states empty, we would want to fill precisely half of the edge states.  As we have argued above, any partial filling of the edge states in a given valley will violate scale invariance.   Thus, if we require scale invariance and charge neutrality, there are two candidate ground states, where the edge states of one valley are completely filled and the edge states of the other valley are completely empty.   These are ``valley ferromagnetic'' states.  It is interesting that we are led to this valley ferromagnetism by symmetry considerations alone, even in the absence of interactions.

There is a third property of many particle states, which might be bothering the reader at this point, and is worthy of comment.  Each single particle state with wave-function $\psi^{(A)}_{k,\ell\pm}(x)$ or $\psi^{(A)}_{0k}(x)$ carries a linear momentum $k$ along the direction parallel to the boundary.  Moreover, as we have already observed, all of the zero mode states in a given valley have the same sign of $k$. The contribution of a given occupied single particle state to the total momentum is
$
k\psi^{(A)\dagger}_{k,\ell\pm}(x)\psi^{(A)}_{k,\ell\pm}(x)
$
or 
$
k\psi^{(A)\dagger}_{0 k}(x)\psi^{(A)}_{0k}(x)
$.
If we add up the contribution of the occupied bulk states in a single valley it is
\begin{align}
&\int_{-\infty}^\infty dk\int_0^\infty d\ell  ~ k~\psi^{(A)\dagger}_{k,\ell-}(x)\psi^{(A)}_{k,\ell-}(x)
=\lim_{x'\to x}\frac{1}{i}\frac{d}{d{x}_2}\int_{-\infty}^\infty dk\int_0^\infty d\ell  \psi^{(A)\dagger}_{k,\ell-}(x')\psi^{(A)}_{k,\ell-}(x)\,.
\end{align}
We can use a particle-hole-symmetry identity, which we find in the next section in equation (\ref{particle_hole_symmetry}),  to re-write the 
right-hand-side of the above equations as
\begin{align}
&\lim_{x'\to x}\frac{1}{i}\frac{d}{dx_2}\int_{-\infty}^\infty dk\int_0^\infty d\ell\; \frac{1}{2}\biggl\{
 \psi^{(A)\dagger}_{k,\ell-}(x')\psi^{(A)}_{k,\ell-}(x)+\psi^{(A)\dagger}_{k,\ell+}(x')\psi^{(A)}_{k,\ell+}(x)
\biggr\}\,.
\end{align}
Then we can use completeness of the set of all wave-functions to rewrite the right-hand-side of the above equation as
\begin{align}
\lim_{x'\to x}\frac{1}{i}\frac{d}{dx_2} \;\frac{1}{2}\biggl\{
\delta(x'-x)-\int dk  \psi^{(A)\dagger}_{0,k}(x')\psi^{(A)}_{0k}(x) 
\biggr\}
 =-\frac{1}{2}\int dk~k~\psi^{(A)\dagger}_{0,k}(x)\psi^{(A)}_{0k}(x)\,, 
\end{align}
where we have taken the $x'\to x$ limit and we have assumed that the delta function
(or some parity symmetric regularization of the delta function) obeys $\lim_{x'\to x}\frac{d}{dx_2}\delta(x-x')=0$. 
We conclude that the bulk states have a deficit of momentum density that can be compensated
by the edge states. In some sense, for a given valley, it should be half of the possible momentum that could be stored in the edge states.

Now, if we add the two valleys together, the deficit of bulk momentum cancels, since the edge state contributions would have opposite signs.  The Dirac vacuum for the bulk states in the two-valley system contributes zero total momentum density.
 Then, the momentum that is left is entirely due to the edge states and our valley ferromagnetic configuration of the edge states carries a
  linear momentum  density in the direction parallel to the boundary.   
  It is also easy find the integrated momentum density, which is a component of the stress tensor as
  \begin{align}
  \left< {\bf T}^{0y}(t,x,y) \right> = \int dk ~k~\psi^{(A)\dagger}_{0k}(x)\psi^{(A)}_{0k}(x) 
  =\pm\frac{1}{4\pi}\frac{1}{x^3}\label{induced_momentum}
  \end{align}
  where the plus of minus sign depends on which valley edge states are filled and which valley is empty. Remember that the integral over $k$ should be over  $(0,\infty)$ for valley $(1)$ and $(-\infty,0)$ for valley $(2)$. This is what leads to the plus or minus sign on the right-hand-side of (\ref{induced_momentum}).
  
  This induced momentum density should characterize our valley ferromagnetic states. 
Its source definitely deserves further exploration.  We can only comment here that the current density, due to charge transport,
by a similar argument to the above is 
 \begin{align}
  \left< J^{y}(t,x,y) \right> =- \int dk ~ ~\psi^{(A)\dagger}_{0k}(x)\gamma^0\gamma^2\psi^{(A)}_{0k}(x) 
  =0 
  \end{align}
  This tells us that the induced momentum density that we found in equation (\ref{induced_momentum}) is not associated with charge transport.  
 
  There is a consequence of this discussion if we consider the total momentum
  $$
  P^y=\int dxdy~{\bf T}^{0y}(t,x,y)
  $$
  and the generator $\Delta$ of scale transformations and the total U(1) charge $Q$.  What we have found is that it is impossible to find
  a distribution of the fermions into edge states so that our candidate ground state obeys all three of the following 
  $$
\Delta |\psi\rangle=0~,~ P^y|\psi\rangle=0~,~Q|\psi\rangle=0
$$
It is possible to have two out of three in this system with two valleys of fermions.

Of course if we further double the fermion degrees of freedom, then it is possible to have all three of the above.  In graphene, where the electrons have spin, this doubling is done for us, it is just the two spin states which to a good approximation are degenerate there.  Once we do this additional doubling of the fermions, we can easily find
states where all three of the above hold, the state is scale invariant, charge neutral and has zero momentum density.  The two possibilities are
\begin{enumerate}
\item{}Both valleys of one spin polarization are filled and both valleys of the other spin polarization are empty.  This is a spin ferromagnetic state.
\item{}For one spin polarization, valley $(1)$ is filled and valley $(2)$ is empty and for the other spin state, valley $(2)$ is filled and valley $(1)$ is empty.  This is a valley, rather than spin ferromagnet.  
\end{enumerate}

These are the maximally symmetric states and they are ferromagnetic.  It is interesting that the valley ferromagnet competes with the spin ferromagnet.  This is a prediction of ours.  In condensed matter studies it is usually found that the lowest energy state is the spin ferromagnet.  Those studies are usually done at the level of the tight-binding lattice theory.  The fact that our continuum theory sees an extra state, the valley ferromagnet,  is attributable to the fact that the extra state should be there in the lattice theory too, but perhaps with a small energy, one which scales to zero in the continuum limit.  In that case, our prediction would be that, as well as the spin ferromagnet which is widely predicted, there is a competing, metastable valley ferromagnet.  

In the next section, we will study the effect of introducing a weak repulsive interaction for the fermions. We will see that our two ferromagnetic states, which in this section we have identified by symmetry arguments alone, survive there and are picked out by the interaction as the true ground states.  

\section{Resolution of the ground state degeneracy}\label{resolution}

Now we must address the question as to why we are so interested in states which are scale invariant. Of course, at the level of the non-interacting theory, there is little that distinguishes scale invariant or charge neutral states from other states that they are degenerate with.   The idea here is that, given the high degree of degeneracy, the introduction of even an infinitesimally weak interaction could split the degeneracy and favour various states and often these are states which have a higher degree of symmetry.   We could, for example, introduce an instantaneous Coulomb interaction, 
\begin{align}\label{interaction_hamiltonian}
H_I= \int d^2xd^2x' ~\rho(x)~\frac{e^2}{8\pi\epsilon_0 |\vec x-\vec x'|} ~\rho(x')\,.
\end{align}
The fact that it is the Coulomb interaction is not overly important to us.  Our arguments in the following would be valid for any density-density interaction
which is repulsive at all scales, translation invariant and has a positive semi-definite interaction Hamiltonian which depends quadratically on the fermion density as 
(\ref{interaction_hamiltonian}) does. 
In the expression  (\ref{interaction_hamiltonian}),  $\rho(x)$ is the (suitably operator ordered) density operator
\begin{align}\label{charge_density}
\rho(x)=\frac{1}{2}\left[ \psi^{(1)\dagger}(x),\psi^{(1)}(x)\right]~+~  \frac{1}{2}\left[\psi^{(2)\dagger}(x) ,\psi^{(2)}(x)\right] \,.
\end{align}
 Since the boundary conditions already break Lorentz invariance we have no incentive to make the interaction Lorentz invariant.  In fact, in a realistic Dirac material, the
 Coulomb interaction is often the most important one.  Moreover, the speed of the Dirac fermions is so much smaller than the speed of light that the electromagnetic interactions are often well approximated by the instantaneous Coulomb interaction contained in (\ref{interaction_hamiltonian}). 
  
 By its classical engineering dimension,  the Coulomb interaction is a marginal deformation of our theory -- it contains no dimensionful parameters and it is scale invariant at the classical level.  However, it is also well known that, for this theory without the boundary, quantum corrections in the form of renormalization of the massless Dirac theory with this Coulomb interaction added violate scale invariance in such a way that interaction becomes a marginally irrelevant perturbation \cite{Sheehy2007,Son2007}.  
 That means that the weak coupling limit is stable under the renormalization group flow to the infrared and the fixed point is represented by the scale invariant free field theory that we have been discussing.   We expect that the Coulomb interaction introduced to this theory with a boundary behaves in a similar way
 although the fine details have yet to be explored.   Here we will assume that it justifies studying the weak coupling limit in order to resolve the vacuum degeneracy. 
 
 Let us consider an ansatz for the ground state which consists of the bulk states having their negative energy levels completely filled and their positive energy levels completely empty and, then, some  filling of the edge states. We will study how the Coulomb interaction prefers to fill the edge states. We will denote a state which is such a bulk Dirac vacuum and  some definite filling of edge states by the symbol $|\alpha\rangle$. Without interactions, the energy of the fermion system does not depend on how the edge states are filled.  There is an enormous degeneracy of ground states in that all states $|\alpha\rangle$ have the same leading order energy. 

We shall assume that the interaction in equation (\ref{interaction_hamiltonian}) is sufficiently weak that we can use first order perturbation theory.  In the first order of degenerate perturbation theory, we must 
resolve the ground state degeneracy by diagonalizing the matrix 
\begin{align}\label{matrix_element}
\delta E_{\alpha\alpha'}~=~\langle\alpha|H_I|\alpha'\rangle
\end{align}
where $|\alpha\rangle,|\alpha'\rangle,\ldots $ are amongst the degenerate ground states of the Hamiltonian in the absence of the interaction.

We will not be able to solve this diagonalization problem in general.  However, we will be able to identify the lowest energy states. 
To do this we examine the operation of the interaction Hamiltonian operator on a state with some definite filling of the edge states. The key to understanding this interaction is to realize that the bulk states in both $|\alpha\rangle$ and $|\alpha'\rangle$ are in the Dirac vacuum with all of the negative energy states filled and positive energy states empty.
There are three separate contributions to the  matrix element $\langle\alpha|H_I|\alpha'\rangle$ in equation (\ref{matrix_element}):
\begin{enumerate}
 \item{}The interaction Hamiltonian can create and then re-annihilate a bulk 
fermion-anti-fermion pair.    This process is does not depend on the filling of the edge states and its contribution to the ground state energy is 
\begin{align}
\delta E^{(1)}_{\alpha\alpha'}=\delta E_{\rm bulk}\delta_{\alpha\alpha'}+\ldots
\end{align}
where
\begin{align}
&\delta E_{\rm bulk}=
 \int d^2xd^2x'~\frac{e^2}{8\pi\epsilon_0|x-x'|}\sum_{A=1,2} \int_0^\infty d\ell d\ell' ~\times \nonumber \\ &\times\int_{-\infty}^\infty dkdk' {\rm Tr}\left[  \psi^{(A)}_{k\ell-}(x)\psi^{(A)\dagger}_{k\ell-}(x')~
\psi_{k',\ell'+}^{(A)}(x')\psi_{k',\ell'+}^{(A)\dagger}(x)\right]
 \label{energy_cont_3}
\end{align}
does not depend on the state $|\alpha\rangle$.  This is the usual expression for the Coulomb exchange energy which one would compute for the Dirac ground state if the edge states were absent.  Generally, this quantity scales like the  volume of space and it is ultraviolet divergent. We will assume that it can be defined with a suitable ultraviolet cutoff, the details of which are of no importance to us as this contribution to the 
matrix that we want to study is already diagonal and it is proportional to the unit matrix, so it is of absolutely no help in distinguishing preferred edge states. 

Worthy of note is the fact that the Dirac commutator choice of ordering the operators in the charge density in (\ref{charge_density}) automatically cancels the total
charge density of the bulk states when the bulk state is the Dirac vacuum.  This is the reason why there is no direct interaction accompanying the exchange interaction
in equation (\ref{energy_cont_3}). 

  \item{}The interaction Hamiltonian can create and re-annihilate a bulk particle or a bulk anti-particle.  This will be accompanied by either annihilation and then creation or creation and then annihilation of  edge states, respectively. The latter processes can happen only when the edge state is suitably occupied or unoccupied. 
  The matrix elements for the  two possible processes are
\begin{align}
& \int d^2xd^2x'~\frac{e^2}{8\pi\epsilon_0|x-x'|} \biggl\{
\sum_{\substack{ k,\ell  \\ A,k'~\rm occupied}}{\rm Tr}\left[  \psi^{(A)}_{k\ell-}(x)\psi^{(A)\dagger}_{k\ell-}(x')~
\psi_{0,k'}^{(A)}(x')\psi_{0,k'}^{(A)\dagger}(x)\right]
 \nonumber \\
&+\sum_{\substack{ k,\ell  \\ A,k'~\rm unoccupied}}{\rm Tr} \left[\psi^{(A)}_{k\ell+}(x)\psi^{(A)\dagger}_{k\ell+}(x')~
\psi_{0,k'}^{(A)}(x')\psi_{0,k'}^{(A)\dagger}(x)\right]
\biggr\}
\label{energy_cont_3}\,,
\end{align}
where we have taken into account the fact that the particle or hole which are created and then destroyed must have the same $k,\ell$ quantum numbers and then, that conservation of total $x_2$-direction momentum requires that the $k'$'s on the edge state wavefunctions are identical.  The summation signs stand for integrals over $\ell,k,k'$ and the bulk wavefunctions in the first term have negative energies whereas those in the second term have positive energies.

It is easy to see from the explicit solutions for the wave-functions that 
the summations over negative and positive energy bulk states are related by 
\begin{align}
& \psi^{(A)}_{k\ell-}(x)\psi^{(A)\dagger}_{k\ell-}(x')= \left[ \begin{matrix} 0 & -i \cr i & 0 \cr \end{matrix}\right] ~ ~\psi^{(A)}_{k\ell+}(x)\psi^{(A)\dagger}_{k\ell+}(x')~~\left[\begin{matrix} 0 & -i \cr i & 0 \cr \end{matrix}\right]
\label{particle_hole_symmetry}
\end{align}
for any $k,\ell$ and that the matrix $\left[\begin{matrix} 0 & -i \cr i & 0 \cr \end{matrix}\right]$ commutes with $\psi_{0,k'}^{(A)}(x')\psi_{0,k'}^{(A)\dagger}(x)$ for any value of $k'$.    This allows us to write equation (\ref{energy_cont_3}) as
\begin{align}
& \int d^2xd^2x' \frac{e^2}{4\pi\epsilon_0|x-x'|}\;
\sum_{\substack{ k,\ell \\ A,k' }}{\rm Tr}\psi^{(A)}_{k\ell+}(x)\psi^{(A)\dagger}_{k\ell+}(x')
\psi_{0,k'}^{(A)}(x')\psi_{0,k'}^{(A)\dagger}(x)
\label{energy_cont_3_1}
\end{align}
where the sum over $k'$ is now a sum over all of the edge states in valley $A$ . 
This sum clearly does not depend on the filling of the edge states.  
As a result
 $$
 \delta E^{(2)}_{\alpha\alpha'}=\delta E_{\rm bulk\leftrightarrow \rm edge}\delta_{\alpha\alpha'}+\ldots
 $$
  In this formula, $\delta E_{\rm bulk\leftrightarrow \rm edge}$ does not depend on the states $|\alpha\rangle$ or $|\alpha'\rangle$. 
  This second type of contribution also is insensitive to the structure of the edge states. 

\item{}Finally, there is a term where the interaction Hamiltonian acts only on the edge states.  To find its properties, we could restrict the charge densities $\rho(x)$ to operators which act on the edge states only. In second quantization, we could do this by simply dropping all of the bulk state creation and annihilation operators from the charge densities.  The restricted charge density operator is Hermitian and the interaction Hamiltonian that is made from this restricted operator is still a positive semi-definite Hermitian operator.    Since it is non-negative, a state in which the expectation value of this restricted interaction Hamiltonian is equal to zero must be an eigenstate with zero eigenvalue and it is thus a lowest energy eigenstate of the edge state restricted interaction Hamiltonian.   It is then also a minimal eigenstate of the matrix
(\ref{matrix_element}) and in linear order of perturbation theory it would be the ground state.  This ground state could still have some residual degeneracy as one would expect to happen in a symmetry breaking situation.

Now, it is easy to see that the only possible zero eigenstates of the restricted charge density operator are the states where, for one valley, all of the edge states are full and for the other valley, all of the edge states are empty.   These ground states are overall charge neutral, and the background charge implicit in the commutator ordering prescription (\ref{charge_density}) cancels between the valleys.  The exchange energy is also zero for those states.
This mechanism for finding the lowest energy states  is reminiscent of Hund's rule for the population of degenerate, or nearly degenerate atomic orbitals or arguments very similar to the one that we have given here for quantum Hall ferromagnetism \cite{semenoff2011}.
\end{enumerate}

Thus we see that there are two degenerate valley ferromagnetic ground states at this lowest order in perturbation theory.   They happen to be identical to the charge neutral and scale invariant states that we constructed in the previous section.

   \section{Conclusions}\label{conclusions} 
   
   In conclusion, we have done a survey of possible boundary conditions for the 2+1 dimensional Dirac theory defined on a half-space.  
   We have focused on two interesting cases, the BCFT which is a boundary conformal field theory and the zig-zag theory which is not, but where the second is the unique 
   example that supports edge states.  We went on to discuss this second example, which we called the zig-zag theory, in some detail.   
   
   The  boundary condition of the zig-zag theory violates Lorentz invariance and also the special conformal invariance.  The problem with these symmetries is the action of the spin operator in their generators.  That action is incompatible with the boundary condition.   On the other hand, the zig-zag   does remain a scale invariant theory and a translation invariant theory, assuming that the boundary of the half-plane is scale and translation invariant, of course.  
   
   We  found that, with two species of fermions, the only states of the zig-zag theory which are scale invariant and charge neutral  are valley ferromagnetic states.  
 In addition, we  observed the puzzling fact  that these valley ferromagnetic states carry an anomalous linear momentum density.  We observed that it is impossible to simultaneously have a state which is scale invariant, charge neutral and vanishing momentum density. 
 
 Of course this does become possible if the fermion species are doubled, which is exactly the case for graphene with spin. To a good approximation the low energy electron dynamics there, the dynamics which is described by the Dirac theory, is spin independent and spin can simply be thought of as an additional degeneracy. 
 On our zig-zag theory on the half-plane, this doubling allows us to find scale invariant, charge neutral states which have vanishing momentum density.  Those states are just the edge spin ferromagnetic state which is widely predicted to occur in graphene with a zig-zag edge and another one which is an edge spin neutral valley ferromagnet. 
 The fact that these states are degenerate in the continuum theory implies that they should be almost degenerate at the lattice level where indications are that the spin ferromagnet is favoured.  It would be interesting to understand in more detail whether this  existence of the valley ferromagnet as a state competing with the spin ferromagnet implies a competing metastable state at the lattice level. This could have very interesting consequences for spintronics applications.
   
   Finally, our proof that our edge ferromagnetic states actually minimize the Coulomb interaction justifies our attention to them.  The proof itself has some novel aspects which are interesting could be of use in other applications.  
   
   It could be that, as techniques for producing clean Dirac semi-metals in the laboratory evolve, that some of what we say here would be tested by experiments.  For a spin polarized case with the zigzag boundary condition, for example, the anomalous momentum density should have some physical consequences which have yet to be worked out.
   
   There has recently been significant discussion of boundary conformal field theories in general \cite{Andrei2020} and specifically boundary conformal field theories with fermions \cite{Sato2021} , higher dimensional versions of the  theory that we discuss here.  It would be very interesting to see if our some of our ideas generalize there.  For example, the existence of edge states of defects or edges of higher than just one dimension is easily seen to occur.  For example, it is easy to see that there are boundary conditions for the four dimensional Dirac operator which, like the ones that we have discussed, violate Lorentz and conformal symmetry, but preserve scale invariance and contain edge states very similar to the zigzag theory that we have discussed there. Although no applications to particle physics or cosmology come to mind, for sure three dimensional Weyl semi-metals should have boundaries and this phenomenon would be a possibility there.

   \begin{acknowledgments}
This work is supported in part by the Natural Sciences and Engineering Research Council of Canada (NSERC).
\end{acknowledgments}

\bibliographystyle{jhep}
\bibliography{draft.bib}
 
\end{document}